\documentclass[12pt]{article}

\usepackage[margin=1in]{geometry} 
\usepackage{amsmath,amsthm,amssymb}
\usepackage{textcomp}
\usepackage[dvips,final]{graphicx}         		
\usepackage[greek, english]{babel}                			
\usepackage[hang,bf]{caption}              			
\usepackage{rotating}                     		 		
\usepackage{subfigure}
\usepackage{units}
\usepackage{fancyhdr}
\usepackage{lscape}
\usepackage{amsfonts}
\usepackage{float}
\usepackage{graphicx}
\usepackage{longtable}
\usepackage{fancybox}
\usepackage{setspace} 
\usepackage{units} 
\usepackage{pdfpages}
\usepackage{lscape} 		
\usepackage{pdflscape}

\usepackage{upgreek}

\usepackage{listings}

\begin{document}
	\bibliographystyle{elsarticle-num}
	\title{The steady state partial slip problem for half plane contacts subject to a constant normal load using glide dislocations} 
	\author{H. Andresen$^{\,\text{a,}}$\footnote{Corresponding author: \textit{Tel}.: +44 1865 273811; \newline \indent \indent \textit{E-mail address}: hendrik.andresen@eng.ox.ac.uk (H. Andresen).}$\,\,$, D.A. Hills$^{\,\text{a}}$, M.R. Moore$^{\,\text{b}}$\\ \\
		\scriptsize{$^{\text{a}}$ Department of Engineering Science, University of Oxford, Parks Road, OX1 3PJ Oxford, United Kingdom} \\
			$\,$\scriptsize{$^{\text{b}}$ Mathematical Institute, University of Oxford, Andrew Wiles Building, Radcliffe Observatory Quarter, }\\
		\scriptsize{Woodstock Road, OX2 6GG Oxford, UK}}

	\date{}
	\maketitle
	\begin{center}
		\line(1,0){470}
	\end{center}
	\begin{abstract}
		\footnotesize{	
			A new solution for general half-plane contact problems subject to a constant normal load together with alternating shear loads and tension in the steady state is presented. The method uses a formulation where a displacement correction is made to the fully stuck contact solution. There will be two outer regions of slip and a central permanent stick zone, which is explicitly established. Thereby, the maximum extent of the slip zones is effectively specified. Cases of small and large tension are studied, that is when the direction of slip is the same or opposing at the ends of the contact, respectively. \\
			}

	\end{abstract}
	\begin{center}
		\line(1,0){470}
	\end{center}

\section{Introduction}

\hspace{0.4cm}Fretting fatigue is associated with contact problems
enduring partial slip, that is, where part of the contact is adhered
and part is in a slipping state. The presence of the contact causes
a stress concentration and slipping in the presence of friction damages
the surface, producing conditions ripe for the nucleation of cracks.
We are currently investigating, experimentally, a number of fretting
strength problems in the laboratory, and it is important to have the
same representative conditions in the experiment as those present
in the prototype. This means understanding both the normal contact
problem and the partial slip analysis. The latter is uncoupled from
the first if the contacting bodies may be idealised as half-planes
and they are made from the same material. The first solution to problems
of this kind was the celebrated Cattaneo analysis \cite{Cattaneo_1938}
of a Hertzian contact (in the original paper a three-dimensional problem)
subject to a constant normal load and an increasing shear force. The
problem was later re-examined by Mindlin \cite{Mindlin_1949} in a series
of papers in which he looked at the response of a contact to an oscillatory
shear force, and also began studies of the case where the normal load
varies. Further developments of the Cattaneo-Mindlin process, extending it to half-plane problems
of general form were obtained independently by Ciavarella \cite{Ciavarella_1998}
and J\"ager \cite{Jaeger_1998}. 

\begin{figure}[b!]
	\centering
	\includegraphics[scale=0.8, trim= 0 0 0 0, clip]{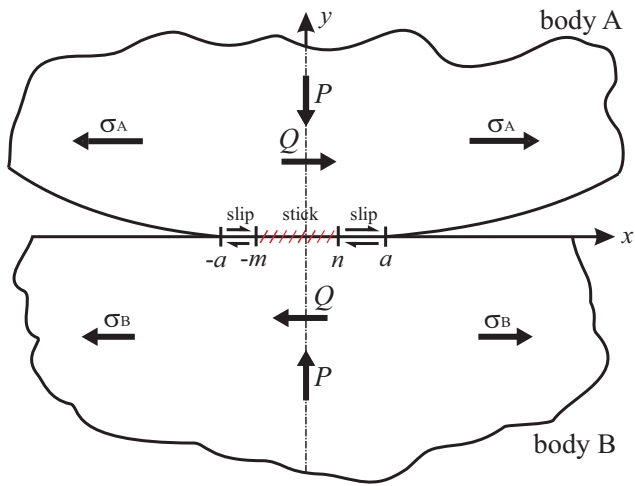}
	\caption{General symmetrical incomplete half-plane contact subject to normal loading and a subsequently applied shear force and bulk tension in partial slip.}
	\label{fig:figure1}	
\end{figure}

\begin{figure}[H]
	\centering
	\includegraphics[scale=0.5, trim= 0 0 0 0, clip]{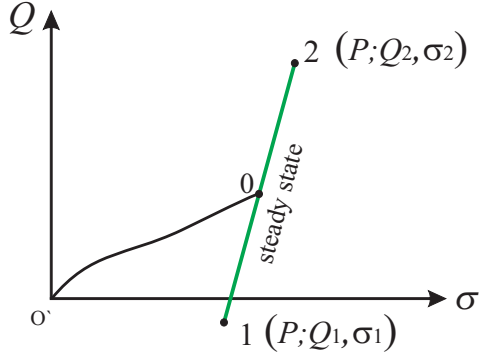}
	\caption{The steady state cycle in $\sigma$-$Q$-space under a constant normal load $P$.}
	\label{fig:figure2}	
\end{figure}

In fact interfacial shear tension may be excited by one of two ways
- either by the action of an external shear force or by the development
of differential tensions between the two bodies, and the first paper
to consider this, prompted by considerations of what arises in a laboratory
fretting test, was by Hills and Nowell \cite{Hills_1987} who solved
the resulting integral equation numerically. A problem of this kind is shown in Figure \ref{fig:figure1}, in which a general symmetrical half-plane contact is subject to a normal load, $P$, and a subsequently applied shear force, $Q$ and differential bulk tension, $\sigma = \sigma_{\text{A}} - \sigma_{\text{B}}$, in partial slip. 
There have also been further considerations of the effects of bulk
tension by Ciavarella and Macina \cite{Ciavarella_2003} and Moore et al. \cite{Moore_2018}. In most practical problems the loads at the contact are excited
by a solitary varying force which does not change orientation. This
means that there is no phase shift between the loads excited, and
a commonly arising problem is when the normal force, $P$, is held
constant and the shear force oscillates between two values ($Q_{1},Q_{2}$),
and, at the same time, the bulk tension fluctuates between
two values ($\sigma_{1},\sigma_{2}$), Figure \ref{fig:figure2}.

Although problems of this kind can be solved by marching-in-time
type methods \cite{Barber_2011}, it is very helpful for the user of these results if the
solution can be obtained quickly and in as near to a closed form as possible.
We have recently obtained a very simple solution to the problem posed
which is appropriate when the applied tension is small, so that the
slip zones at each edge of the contact are of the same sign \cite{Andresen_2019_2}.
The solution developed in that paper is a variant of the Ciavarella-J\"ager approach,
developed further by Barber et al \cite{Barber_2011}, in which the
fundamental form of the shear traction distribution is taken as the
sum of a sliding shear traction together with a corrective term applied
over the stick region $[-m,\qquad n]$, where $m, n >0$, to restore stick there, Figure \ref{fig:figure1}. This
method has stood the test of time well but it is extremely difficult
to adapt it to problems where the slip zones are of opposite sign,
at any rate in a way which can be solved in closed form.

A breakthrough was the discovery of the solution for the shear tractions
induced on the interface by a dislocation present on that interface
between two elastically similar half-planes joined over an interval
-- here {[}$-a,\qquad a${]} see Figure \ref{fig:figure1} \cite{Moore_2018_2}. This meant that the reverse assumptions
might be made compared with those described above, viz. that the contact
was adhered everywhere, and then glide dislocations inserted
over the putative slip regions to restore slip there to its limiting
value, so the shear traction, $q(x)$, in the slip region is given by $q(x)=f p(x)$, where $f$ is the coefficient of friction and $p(x)$ is the normal pressure distribution. This method led to tractable integrals for both small-tension
and large-tension cases, that is when the direction of slip is the same or opposing at the ends of the contact, and a marching-in-time procedure was adopted \cite{Moore_2018}. The object in the present paper is to extend the
solutions in \cite{Andresen_2019_2} to those for a steady state response, and in particular
to the case where the bulk tension may be large. At the same time, in the present paper, the normal load will be kept constant as this permits the integral equations to be developed and the solution to be found in closed-form.

\section{Adhered Condition}

\begin{figure}[H]
	\centering
	\includegraphics[scale=0.445, trim= 0 0 0 0, clip]{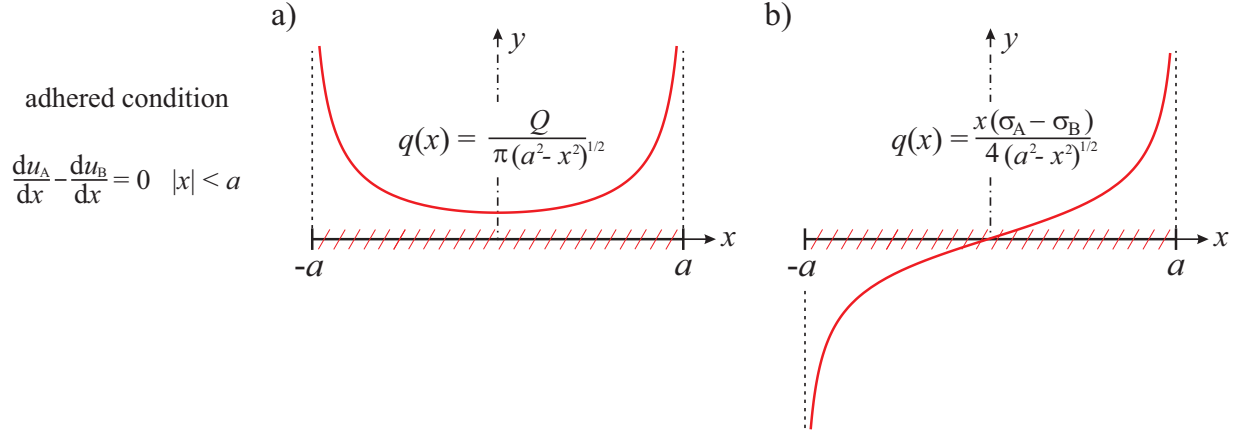}
	\caption{Shear tractions in the fully adhered state due to a) shear force only and b) bulk tension only.}
	\label{fig:figurefullyadhered}	
\end{figure}

\hspace{0.4cm}The starting point for the solution
is one where the contact has already been formed, so that there is
no differential surface strain present, and the contact, associated
with a normal load $P$, occupies the interval {[}$-a,\qquad a${]}.
If the coefficient of friction, $f$, was indefinitely high, all slip would be prevented when a shear force, $Q$, was subsequently applied. Therefore, the difference in strains parallel with the surface between the two bodies will be zero, i.e. $\frac{\mathrm{d}u_{\text{A}}}{\mathrm{d}x} - \frac{\mathrm{d}u_{\text{B}}}{\mathrm{d}x} = 0  \,,|x|<a$, and the shear traction is as shown in Figure \ref{fig:figurefullyadhered} a). Equally, if tensions, $\sigma_A, \sigma_B$, are developed in each body as shown in Figure \ref{fig:figure1} the shear tractions arising will be as shown in Figure \ref{fig:figurefullyadhered} b).

We note that, if the coefficient of friction, $f$, were high enough to prevent
all slip, the change in shear traction, $\Delta q(x)$, when a change
in shear force, $\Delta Q$, and change in bulk tension, $\Delta\sigma$,
are imposed, when we move from load state 1 to load state 2, Figure \ref{fig:figure2}, is given by 
\begin{align}
\Delta q(x)=\frac{\Delta Q}{\pi\sqrt{a^{2}-x^{2}}}+\frac{x\Delta\sigma}{4\sqrt{a^{2}-x^{2}}}.
\end{align}

In this problem the change in shear load is
\begin{align}
\Delta Q=Q_{2}-Q_{1},
\end{align}
and change in bulk tension is
\begin{align}
\Delta\sigma=\sigma_{2}-\sigma_{1},
\end{align}
where $\sigma_1$ and $\sigma_2$ represent the difference in bulk tension\footnote{In general, bulk stresses may arise in each body. When this is the case, providing they are synchronous, we define $\sigma_i=\sigma_{B,i}-\sigma_{A,i}$, for $i=1,2$ representing two load points in the steady state (see Figure \ref{fig:figure2}).} between the two bodies at the ends of the load cycle, respectively. If both $\Delta Q$ and $\Delta\sigma$, are positive, Figure \ref{fig:figure1},
we see that at the right hand edge of the contact their effects will sum (as $x\rightarrow a^{-}$) whilst at the left hand edge, as $x\rightarrow-a^{+}$ their effects tend to cancel.

\section{Partial Slip Steady State Problem}

\hspace{0.4cm}Consider a distribution of glide dislocations, of density $D(x)$,
present over an interval {[}$-m,\qquad n${]} on the junction between
two half-planes over {[}$-a,\qquad a${]}. This will give rise to
a shear traction, $q(x)$, given by\footnote{We interpret these integrals in the usual way: Cauchy principal value when $x$ lies within their interval, otherwise regular.} \cite{Moore_2018} 
\begin{align}
q(x)=\frac{1}{\pi A \sqrt{a^{2}-x^{2}}}\int_{-m}^{n}\frac{\sqrt{a^{2}-\xi^{2}}D\left(\xi\right)}{\xi-x}d\xi\; \text{,} \qquad -a < x < a \; \text{.} 
\end{align}

The contact region is over $-a<x<a$ as shown in Figure \ref{fig:figure1}, and the solution to be developed is mathematically exact only if the materials considered have elastically similar properties so that the material compliance of the two touching bodies is given by \cite{Barber_2010}
\begin{align}
A = \frac{\kappa + 1}{2 \mu}\; \text{,}
\end{align}
where $\mu$ is the shear modulus and $\kappa$ is Kolosov's constant, defined as 
\begin{align}
\kappa=
\begin{cases}
3-4\nu \qquad \; \; \; \; \; \, \text{ for plane strain}\,  \text{,} \\
\displaystyle{\frac{3-\nu}{1+\nu}} \qquad \qquad \text{ for plane stress} \,  \text{,}
\end{cases}
\end{align}
where $\nu$ is Poisson's ratio. Suppose we are in steady state and the loads are at state $1$, see Figure \ref{fig:figure2}, and the increment of loading has just changed sign
 and is now increasing. The shear traction which is locked-in at that point is denoted $q_{1}^{+}(x)$. As the loads approach load state $2$, the shear traction can be written as 
\begin{align}
q_{2}^{-}(x)= & \,q_{1}^{+}(x)+\frac{\Delta Q}{\pi\sqrt{a^{2}-x^{2}}}+\frac{x\Delta\sigma}{4\sqrt{a^{2}-x^{2}}}+\label{point2-_SIE}\nonumber \\
 &\frac{1}{\pi A\sqrt{a^{2}-x^{2}}}\left[\int_{-a}^{-m}\frac{\sqrt{a^{2}-\xi^{2}}D^{L}\left(\xi\right)}{\xi-x}d\xi+\int_{n}^{a}\frac{\sqrt{a^{2}-\xi^{2}}D^{L}\left(\xi\right)}{\xi-x}d\xi\right]\,,
\end{align}
where $x\in(-a,a)$ and $D^L(x)$ is the density of dislocations arising during the loading phase and $D^U(x)$ is the density of dislocations arising during the unloading phase. When we `move around the corner' at load state $2$ this shear traction,
in turn, becomes locked-in and is denoted $q_{2}^{+}(x)$. Now, as we approach point $1$ we can write
\begin{align}
q_{1}^{-}(x)= & \,q_{2}^{+}(x)-\frac{\Delta Q}{\pi\sqrt{a^{2}-x^{2}}}-\frac{x\Delta\sigma}{4\sqrt{a^{2}-x^{2}}}+ \label{point1-_SIE}\nonumber \\
 & \frac{1}{\pi A\sqrt{a^{2}-x^{2}}}\left[\int_{-a}^{-m}\frac{\sqrt{a^{2}-\xi^{2}}D^{U}\left(\xi\right)}{\xi-x}d\xi+\int_{n}^{a}\frac{\sqrt{a^{2}-\xi^{2}}D^{U}\left(\xi\right)}{\xi-x}d\xi\right]\,, 
\end{align}
where $x\in(-a,a)$. In each equation, the glide dislocation distributions over
the slip intervals physically represent the slip displacements there,
and the shear tractions vanish outside the intervals stated.

\begin{figure}[H]
	\centering
	\includegraphics[scale=0.39, trim= 0 0 0 0, clip]{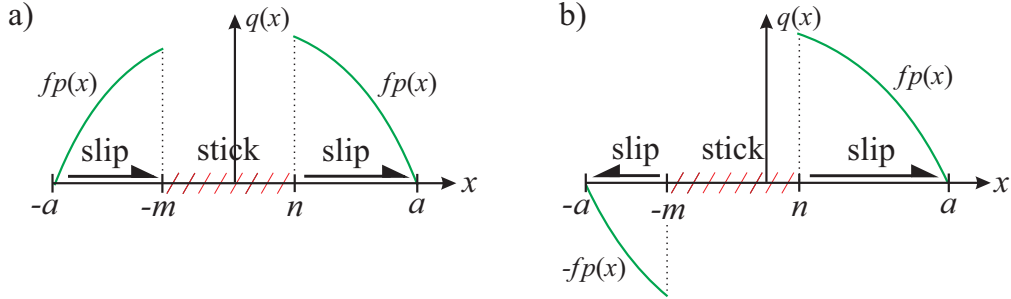}
	\caption{Shear tractions for; (a) the small-tension case, (b) the large-tension case in partial slip.}
	\label{fig:large_small_tension_case}	
\end{figure}

If we subtract integral \eqref{point1-_SIE}
from integral \eqref{point2-_SIE} we arrive at the following integral 
\begin{align} \label{eqpoint2-eqpoint1}
q_{2}^{-}(x) - q_{1}^{-}(x) = & \,q_{1}^{+}(x) - q_{2}^{+}(x) + \frac{2 \Delta Q}{\pi\sqrt{a^{2}-x^{2}}}+ \frac{2 x\Delta\sigma}{4\sqrt{a^{2}-x^{2}}} + \nonumber\\
&\frac{1}{\pi A\sqrt{a^{2}-x^{2}}}\left[\int_{-a}^{-m}\frac{\sqrt{a^{2}-\xi^{2}}D^{L}\left(\xi\right)}{\xi-x}d\xi+\int_{n}^{a}\frac{\sqrt{a^{2}-\xi^{2}}D^{L}\left(\xi\right)}{\xi-x}d\xi -\right. \nonumber \\
&  \left. \qquad \qquad \qquad \quad \int_{-a}^{-m}\frac{\sqrt{a^{2}-\xi^{2}}D^{U}\left(\xi\right)}{\xi-x}d\xi-\int_{n}^{a}\frac{\sqrt{a^{2}-\xi^{2}}D^{U}\left(\xi\right)}{\xi-x}d\xi\right], 
\end{align}
where $x\in(-a,a)$. We know that, in the slip regions, the shear tractions are equal to
their limiting value, i.e. 
\begin{align}
q_{2}^{-}(x)=q_{2}^{+}(x)=fp(x)\,\text{sgn}(x)\qquad-a\leq x\leq-m\qquad\mbox{and}\qquad n\leq x\leq a
\end{align}
and 
\begin{align}
q_{1}^{-}(x)=q_{1}^{+}(x)=-fp(x)\,\text{sgn}(x)\qquad-a\leq x\leq-m\qquad\mbox{and}\qquad n\leq x\leq a,
\end{align}
where we should \emph{omit} the $\text{sgn}(\bullet)$ functions
when we are considering the small-tension case, Figure \ref{fig:large_small_tension_case}(a), and \emph{include} them when we are considering the large-tension
case, Figure \ref{fig:large_small_tension_case}(b).

We have already used the principle of conservation of material in
assuming that the maximum extent of slip penetration is the same at
points $1^{-},2^{-},$ but we can also say that the change in the
slip displacement of all opposing particles within the slip zones during the
loading and unloading phases is equal in magnitude and opposite in sign. For example,
in the right hand slip region we see that the slip
displacement at point $x$, $u^L(x)$, during loading is given by 
\begin{align}
u^{L}(x)=\int_{n}^{x}D^{L}\left(\xi\right)d\xi\qquad n<x<a.
\end{align}

We know that during unloading we will recover the slip displacement,
i.e. $u^{L}(x)+u^{U}(x)=0$ within the slip zones. Hence, we infer that 
\begin{align}
D^{U}(x)=-D^{L}(x)\qquad-a\leq x\leq-m\qquad\mbox{and}\qquad n\leq x\leq a.
\end{align}

If we make use of these results and restate equation \eqref{eqpoint2-eqpoint1} for the slip zones only, we arrive at the following integral equation, in terms of the dislocation
density $D^{L}(x)$
\begin{align}\label{eq13}
2fp  &(x)\,\text{sgn}(x)-\left(\frac{\Delta Q}{\pi\sqrt{a^{2}-x^{2}}}+\frac{x\Delta\sigma}{4\sqrt{a^{2}-x^{2}}}\right)= \nonumber\\
  =&\frac{1}{\pi A\sqrt{a^{2}-x^{2}}}\left[\int_{-a}^{-m}\frac{\sqrt{a^{2}-\xi^{2}}D^{L}\left(\xi\right)}{\xi-x}d\xi+\int_{n}^{a}\frac{\sqrt{a^{2}-\xi^{2}}D^{L}\left(\xi\right)}{\xi-x}d\xi\right],
\end{align}
where $x\in(-a,-m)\cup(n,a)$.

\section{Solution}

\hspace{0.4cm}The formulation works equally well with the same or opposite signs
of slip, so we start by making some progress towards a general solution. We note that the left-hand side of equation
\eqref{eq13} may be interpreted as the difference between the fully stuck and
slipping shear traction distributions within the slip region, which
we will denote by $q_{c}(x)$, and hence we may write 

\begin{align} \label{corrective_shear_traction}
q_{c}(x) = & -\frac{\Delta Q}{\pi \sqrt{a^{2}-x^{2}}}-\frac{\Delta\sigma x}{4 \sqrt{a^{2}-x^{2}}}+ 2 f p(x) \,\text{sgn}(x) = \nonumber\\
 = &\frac{1}{\pi A \sqrt{a^{2}-x^{2}}}\left[\int_{-a}^{-m}\frac{\sqrt{a^{2}-\xi^{2}}D^{L}(\xi)}{\xi-x}\,\mbox{d}\xi+\int_{n}^{a}\frac{\sqrt{a^{2}-\xi^{2}}D^{L}(\xi)}{\xi-x}\,\mbox{d}\xi\right],
\end{align}
where $x\in(-a,-m)\cup(n,a)$. 
A bounded-both-ends solution for equation \eqref{corrective_shear_traction} is sought, and inverting this equation in the standard way leads to the two consistency conditions \cite{Moore_2018}
\begin{align}
0=\int_{-a}^{-m}\frac{q_{c}(\xi)\xi^{k}}{\sqrt{(\xi-n)(\xi+m)}}\,\mbox{d}\xi-\int_{n}^{a}\frac{q_{c}(\xi)\xi^{k}}{\sqrt{(\xi-n)(\xi+m)}}\,\mbox{d}\xi ,
\end{align}
where $k=0,1$. Making use of the knowledge that 
\begin{align}
\int_{-a}^{-m}\frac{\xi^{k}}{\sqrt{a^{2}-\xi^{2}}\sqrt{(\xi-n)(\xi+m)}}\,\mbox{d}\xi=\begin{cases}
0 & \mbox{for}\;k=0,\\
-\pi & \mbox{for}\;k=1,\\
{\displaystyle {\frac{(m-n)\pi}{2}}} & \mbox{for}\;k=2,
\end{cases}
\end{align}
we can reduce the consistency conditions to the following pair of
equations for $m$, $n$
\begin{align}
\frac{\Delta\sigma\pi}{8}= & \,-\int_{-a}^{-m}\frac{fp(\xi)\,\text{sgn}(\xi)}{\sqrt{(\xi-n)(\xi+m)}}\,\mbox{d}\xi+\int_{n}^{a}\frac{fp(\xi)\,\text{sgn}(\xi)}{\sqrt{(\xi-n)(\xi+m)}}\,\mbox{d}\xi,\label{eq17}\\
\frac{\Delta Q}{2}+\frac{(n-m)\Delta\sigma\pi}{16}= & \,-\int_{-a}^{-m}\frac{\xi fp(\xi)\,\text{sgn}(\xi)}{\sqrt{(\xi-n)(\xi+m)}}\,\mbox{d}\xi+\int_{n}^{a}\frac{\xi fp(\xi)\,\text{sgn}(\xi)}{\sqrt{(\xi-n)(\xi+m)}}\,\mbox{d}\xi.\label{eq18}
\end{align}

The LHS of equation \eqref{eq17} is constant and independent of the contact geometry. The RHS is specific to the geometry and applies whether small or large tension is being considered, i.e $\text{sgn}\left(\bullet\right)$ is \textit{included} or \textit{omitted}. Similarly, the LHS of equation \eqref{eq18} is geometry-independent, whereas the RHS is geometry and tension case-dependent.

\subsection{Small Tension - Hertz Example}
\hspace{0.4cm} We start by re-evaluating the small-tension case,
for which a solution already exists based on the Ciavarella-J\"ager
prinicple \cite{Andresen_2019_2}. As the slip direction will be the same at both ends of the contact, Figure \ref{fig:large_small_tension_case} (a), the $\text{sgn}\left(\bullet\right)$ function is \textit{omitted} from equations \eqref{eq17} and \eqref{eq18}. Consider a Hertzian contact, where the pressure
distribution is semi-elliptical in form, and may be written as 
\begin{align}\label{Hertz_pressure}
p(x)=\frac{1}{A R}\sqrt{a^2-x^{2}}, \qquad |x|\leq a.
\end{align}

So, by making use of the following standard results \cite{Moore_2018}
\begin{align}
\int_{-a}^{-m}\frac{\xi^{k}\sqrt{a^{2}-\xi^{2}}}{\sqrt{(\xi-n)(\xi+m)}}\,\mbox{d}\xi=\begin{cases}
{\displaystyle {\frac{(n-m)\pi}{2}}} & \mbox{for}\;k=0,\\
{\displaystyle {\left(\frac{(n-m)^{2}}{4}+\frac{(n+m)^{2}}{8}-\frac{a^{2}}{2}\right)\pi}} & \mbox{for}\;k=1,
\end{cases}
\end{align}
the first consistency condition, equation \eqref{eq17}, provides the result that 
\begin{align}\label{eq21}
\frac{(n-m)}{2}=-\frac{A R\Delta \sigma }{8 f},
\end{align}
and after a little manipulation, the second consistency condition, equation \eqref{eq18},
gives 
\begin{align}\label{eq22}
\left(\frac{(n+m)}{2}\right)^{2}=a^{2}-\frac{A R \text{$\Delta $Q}}{\pi  f}.
\end{align}

These are the same results as those obtained by \cite{Andresen_2019_2}. Provided $\Delta Q$ and $\Delta\sigma$ are positive, the same sign slip direction is maintained as long as the change in bulk stress satisfies
\begin{align}\label{inequality}
\Delta\sigma \leq \frac{8 f}{A R} \left(a-\sqrt{a^2-\frac{A R \text{$\Delta $Q} }{\pi  f}}\right)\;.
\end{align}

\subsection{Large Tension - Hertz Example}

\hspace{0.4cm}Suppose now, that the change in tension is large enough for inequality \eqref{inequality} to be violated so that the slip directions are of opposite sign as shown in Figure \ref{fig:large_small_tension_case}(b). The $\text{sgn}\left(\bullet\right)$ function is therefore now \textit{included} in equations \eqref{eq17} and \eqref{eq18}. In a Hertzian example, where equation \eqref{Hertz_pressure} again applies, we use the following results to evaluate the consistency conditions \cite{Moore_2018}
\begin{eqnarray}
	& & \int_{-a}^{-m}\frac{s^{j-1}\sqrt{a^{2}-s^{2}}}{\sqrt{(s-n)(s+m)}}\,\mbox{d}s + \int_{n}^{a}\frac{s^{j-1}\sqrt{a^{2}-s^{2}}}{\sqrt{(s-n)(s+m)}}\,\mbox{d}s = \nonumber \\
	& & \quad = \alpha_{j}\mbox{E}(\chi) + \beta_{j}\mbox{K}(\chi) + \gamma_{j}\Pi\left(\frac{m-a}{m+a},\chi\right) + \delta_{j}\Pi\left(\frac{a-n}{a+m},\chi\right), 
\end{eqnarray}
for $j=1,2$ and where $\mbox{K}(\chi)$, $\mbox{E}(\chi)$, and $\Pi(t,\chi)$ are elliptic integrals of the first, second and third kind and
\begin{alignat*}{5}
\alpha_{1} & = && -2\sqrt{(a+m)(a+n)}, && \; \alpha_{2} && = && \frac{3(m-n)}{2}\sqrt{(a+m)(a+n)}, \\ 
\beta_{1} & = && (n-m+4a)\sqrt{\frac{a+m}{a+n}}, && \; \beta_{2} && = && \frac{-3}{2}\sqrt{\frac{a+m}{a+n}}\left(\left(\frac{n}{3}+2a\right)m-\frac{1}{2}(m^{2}+n^{2})-\frac{2a}{3}(a+n)\right), \\
\gamma_{1} & = && \frac{2a(n-m)}{\sqrt{(a+m)(a+n)}}, && \; \gamma_{2} && = && \frac{3a}{2\sqrt{(a+m)(a+n)}}\left(n^{2}-\frac{2mn}{3}+m^{2}-\frac{4a^{2}}{3}\right), \\
\delta_{1} & = && \frac{m^{2}-n^{2}}{\sqrt{(a+m)(a+n)}}, && \; \delta_{2} && = && \frac{-3(m+n)}{4\sqrt{(a+m)(a+n)}}\left(n^{2}-\frac{2mn}{3}+m^{2}-\frac{4a^{2}}{3}\right), \,\text{and}\\
\chi & = && \frac{\sqrt{(a^{2}-m^{2})(a^{2}-n^{2})}}{(a+m)(a+n)}. && && &&
\end{alignat*}

Therefore, $m$ and $n$ must satisfy the following nonlinear equations
\begin{align}\label{eq25}
	\frac{\Delta\sigma\pi}{8} =&\frac{f}{A R}\left[\alpha_{1}\mbox{E}(\chi)+\beta_{1}\mbox{K}(\chi) + \gamma_{1}\Pi\left(\phi,\chi\right) + \delta_{1}\Pi\left(\omega,\chi\right)\right],
\end{align}
\begin{align}\label{eq26}	
	\frac{\Delta Q}{2} + \frac{(n-m)\Delta\sigma\pi}{16}=&\frac{f}{A R}\left[\alpha_{2}\mbox{E}(\chi)+\beta_{2}\mbox{K}(\chi) + \gamma_{2}\Pi\left(\phi,\chi\right) + \delta_{2}\Pi\left(\omega,\chi\right)\right],
\end{align}
where $\phi=\frac{m-a}{m+a}$ and $\omega = \frac{a-n}{a+m}$.
Equations \eqref{eq25} and \eqref{eq26} are sufficient to define the size of the stick zone for given numerical values of $\Delta Q$ and $\Delta\sigma$.

\subsection{Display of Example Problem Results}
\hspace{0.4cm}
The results presented in the above sections show that the solution behaviour of a Hertzian contact is fundamentally different for small and large tension. In the case of small tension, we arrive at a set of two uncoupled explicit equations, eq. \eqref{eq21} and \eqref{eq22}. In terms of  tangential load, the \textit{extent} of the permanent stick zone is solely dependent on the change in shear force, $\Delta Q$. Its \textit{position}, or eccentricity, on the other hand, is solely dependent on the change in bulk stress, $\Delta\sigma$. In the case of large tension, the two consistency conditions are nonlinearly coupled, and they must be treated numerically in order to determine the extent and position of the permanent stick zone.

\begin{figure}[H]
	\centering
	\includegraphics[scale=0.6, trim= 0 0 0 0, clip]{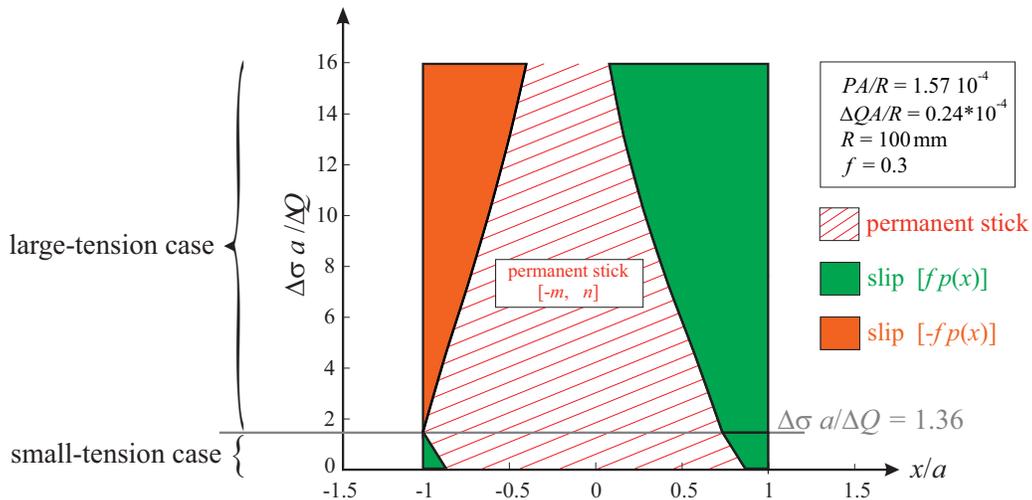}
	\caption{Steady state solutions of maximum slip and permanent stick zone extents for different values of a change in bulk stress, $\Delta\sigma$.}
	\label{fig:evolution_of_slipstick_with_DeltaSigma}	
\end{figure}

Figure \ref{fig:evolution_of_slipstick_with_DeltaSigma} displays \textit{steady state} solutions, i.e. the permanent stick zone and maximum extent of the slip zones, for a Hertzian geometry. The contact is subject to a constant normal load, $P$. The shear load fluctuation, $\Delta Q$, is also kept constant. The permanent stick zone, spanning $[-m, \quad n]$, is then given for different values of $\Delta\sigma\, a/\Delta Q$. Here, we distinguish between the small and large-tension case. We see that, as the change in bulk stress is increased, the extent of the permanent stick zone is indeed constant and its position shifts linearly towards the left-hand contact edge and the direction of slip is the same at both ends of the contact. When the change in bulk stress is high enough so that inequality \eqref{inequality} is violated, the other branch of the solution becomes valid. From this point onwards the behaviour of the solution is highly nonlinear. The extent and position of the permanent stick zone are coupled and the directions of slip oppose at either end of the contact.

\section{Conclusions}
\hspace{0.4cm}The paper provides a method for finding the size of the permanent stick zone for the problem of
a general half plane contact subject to a constant normal load together with alternating shear loads and tension in the steady state. The solution is appropriate both when the tension is small and large, that is the direction of slip is the same or opposing at the ends of the contact, respectively.
Progress is made towards a general solution, which serves as the basis for obtaining steady state solutions for geometries capable of representation within half-plane theory. Closed-form solutions for the small and large-tension case are given for a Hertzian geometry. The difference in behaviour of the solution branches is described and displayed for a Hertzian example. This provides insight into how the maximum extent of the slip zones is affected by the change in differential bulk tension during the steady state.

\section*{Acknowledgements}
\hspace{0.4cm}This project has received funding from the European Union's Horizon 2020 research and innovation programme under the Marie Sklodowska-Curie agreement No 721865. David Hills thanks Rolls-Royce plc and the EPSRC for the support under the Prosperity Partnership Grant “Cornerstone: Mechanical Engineering Science to Enable Aero Propulsion Futures”, Grant Ref: EP/R004951/1.

\bibliography{Contribs_References_Hendrik_Jan2019}

\end{document}